\documentclass[10pt]{llncs}
\usepackage[margin=1in]{geometry}
\usepackage{float}
\usepackage{graphicx}
\usepackage[utf8]{inputenc}
\usepackage{amssymb}
\usepackage{svg} 
\usepackage[linesnumbered,ruled]{algorithm2e}
\usepackage{amsmath}
\usepackage[sorting=none]{biblatex}
\begin{document}

\title{scBeacon: single-cell biomarker extraction via identifying paired cell clusters across biological conditions with contrastive siamese networks}
%
%
\author{Chenyu Liu\inst{1,2}\orcidID{0009-0002-0733-5471} \and
Yong Jin Kweon\inst{2}\orcidID{0000-0001-6972-0063} \and
Jun Ding\inst{2*}\orcidID{0000-0001-5183-6885}}
\authorrunning{F. Author et al.}
%
\institute{School of Computing, National University of Singapore, Singapore\\
\and
Research Institute of the McGill University Health Centre, Department of Medicine, Meakins-Christie Laboratories, McGill University, Montreal, Quebec, H4A 3J1, Canada\\
 *All correspondence should be addressed to JD (jun.ding@mcgill.ca)}

\maketitle              
\begin{abstract}
Despite the breakthroughs in biomarker discovery facilitated by differential gene analysis, challenges remain, particularly at the single-cell level. Traditional single-cell biomarker methodologies heavily rely on user-supplied cell annotations to identify the differentially expressed genes as candidate biomarkers for each of the cell types. However, it is often very challenging to annotate cell types accurately, particularly across different disease conditions that may shift the expression profiles of the same cell types significantly, which increases the difficulty in effectively identifying differential genes between conditions, such as healthy versus disease states, for various cell types. In response, here we introduce scBeacon, an innovative framework built upon a deep contrastive siamese network. scBeacon pioneers an unsupervised approach, adeptly identifying matched cell populations across varied conditions, enabling a refined differential gene analysis. By utilizing a VQ-VAE framework, a contrastive siamese network, and a greedy iterative strategy, scBeacon not only effectively pinpoints differential genes but also identifies the same cell population across different biological conditions, essentially forming cluster pairs. Comprehensive evaluations on a diverse array of datasets validate scBeacon's superiority over existing single-cell differential gene analysis tools. Its precision and adaptability underscore its significant role in enhancing diagnostic accuracy in biomarker discovery. With the emphasis on the importance of biomarkers in diagnosis, scBeacon is positioned to be a pivotal asset in the evolution of personalized medicine and targeted treatments powered by single-cell genomics.
\end{abstract}

\pagebreak
\section{INTRODUCTION}

Conventional biomarker discovery, particularly in bulk analysis, has primarily revolved around the examination of differential gene expression, with a focus on differentially expressed genes (DEGs)~\cite{sanchez2022improved, li2021bulk, meng2023comprehensive}. This approach has yielded valuable insights into transcriptomic variations and their implications across a wide range of diseases~\cite{van2022clinical, zhang2021machine}. The advent of single-cell RNA sequencing has ushered in an unprecedented level of resolution, offering the promise of a more nuanced understanding of biomarkers relevant to various complex diseases~\cite{yue2023introduction}. However, despite these advancements, navigating differential gene analysis at the single-cell level remains fraught with challenges~\cite{das2022differential, van2020trajectory}.

In the realm of single-cell RNA sequencing (scRNA-seq) analysis, a prominent challenge lies in the intricate task of aligning cell clusters across diverse biological conditions. Establishing connections between cells of the same type or subtype in both healthy and diseased conditions for the purpose of differential gene expression analysis poses a non-trivial endeavor. While this alignment is relatively less complex at the bulk level, it becomes considerably more intricate with the granularity introduced by single-cell data~\cite{love2014moderated, law2014voom}. Conventional methodologies heavily rely on user-provided cell annotations. Following single-cell clustering analysis, researchers manually assign cell type information to these clusters, typically guided by established biomarkers. Alternatively, some methods opt to co-cluster cells from different conditions to define cell clusters (cell types or subtypes) and subsequently segregate cells from distinct conditions within these clusters for differential gene analysis. However, such approaches often struggle to pair corresponding cells across conditions due to substantial shifts in cellular states induced by diseases, hindering straightforward cluster pair establishment through naive cell co-clustering. Among the existing methods, DEsingle stands out for its ability to detect differentially expressed genes (DEGs) by modeling the unique distributional characteristics of scRNA-seq data~\cite{klein2015droplet}. Conversely, MAST offers versatility by providing a framework that accommodates both discrete and continuous cellular states within the context of differential expression analysis~\cite{zhang2019enimpute}. scVI is another noteworthy tool that employs a generative model to address challenges related to dropouts and capture the variations inherent in scRNA-seq data~\cite{ye2019decent}. However, despite the significant strides made in advancing scRNA-seq analysis, these methods typically examine expression data independently for each condition, often overlooking the critical interplay between conditions—an essential element in the identification of DEGs within the broader context of understanding differences between biological conditions (e.g., healthy vs. diseased).

Given the formidable challenges presented by current methodologies, there is a pressing need for an automated, unsupervised approach capable of adeptly identifying matched cell populations across different conditions, thus facilitating a more precise analysis of differential gene expression. In response to this imperative, we introduce \textit{scBeacon}, an innovative framework founded upon a deep contrastive siamese network. The primary objective of \textit{scBeacon} is to discern cluster pairs between two conditions and subsequently zero in on the differential genes, which could emerge as pivotal biomarkers. \textit{scBeacon} operates in a three-fold manner: Initially, it leverages a VQ-VAE framework to craft an optimal embedding that effectively represents cells in each biological condition. Subsequently, employing a contrastive siamese network, enhanced by a resourceful iterative strategy~\cite{chen2020simple}, it meticulously identifies the best cluster pairs spanning conditions. Finally, \textit{scBeacon} isolates the differential genes from these automatically discerned cell clusters, spotlighting potential biomarkers. This innovative framework liberates differential gene analysis from the constraints of cell type annotations, whether they are manual or automated, or naive cell co-clustering, championing a more authentic, unsupervised methodology for the inference of cell cluster pairs between biological conditions and the downstream biomarker discovery.

Through the extensive application of \textit{scBeacon} on a gamut of real and simulated datasets, empirical evidence underscores its supremacy over benchmarked, state-of-the-art single-cell differential gene analysis methods. Its precision, coupled with its adaptability, accentuates its potential to herald groundbreaking strides in disease diagnosis, particularly within the domain of single-cell biomarker discovery. As the medical community grapples with multifaceted diseases, tools like \textit{scBeacon}, that amplify our understanding at an intricate level, stand poised to redefine the landscape of personalized medicine and targeted therapeutic interventions.

\section{METHODS}
We present an unsupervised deep learning framework that refines single-cell data representations through iterative contrastive learning, employing a siamese network architecture. The process begins with a neural network encoder that transforms the cell count matrix into lower-dimensional embeddings. These embeddings are then matched to the nearest cluster centroids during the VQ-VAE phase, which is critical for the training of the twin auto-encoder—the central element of the siamese network. In the contrastive learning step, cell pairs are formed based on centroids that have been paired across different conditions; those from the same centroid pairing are considered positive, while those from different ones are negative. These pairs are then processed by the twin auto-encoder to obtain cell embeddings that are conducive to the contrastive learning task. By fine-tuning the alignment of positive cell pairs against negative pairs in the latent space, we iteratively refine the cell embeddings, integrating the losses from VQ-VAE and contrastive learning. This results in continuously enhanced cell embeddings, which lead to updated clustering and establish a feedback loop for ongoing refinement, persisting until a convergence criterion is met. After determining the cell cluster pairs, we proceed to pinpoint differentially expressed genes between these paired clusters, which serve as candidate biomarkers for distinguishing the specific cell types under varying conditions. 

\begin{figure}[!htb]
\includegraphics[width=\textwidth]{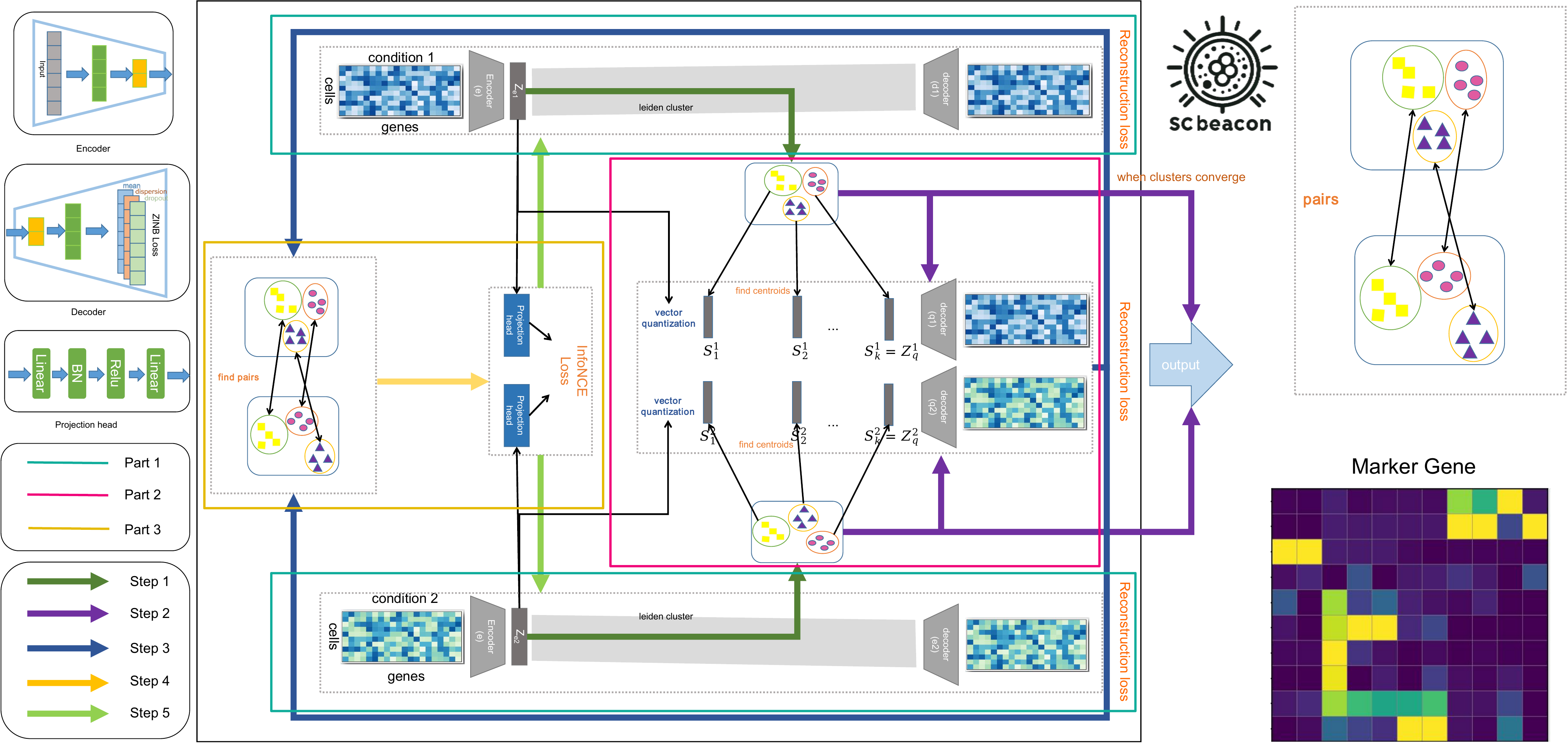}
\caption{\textbf{scBeacon Framework Overview:} ~\textcolor{blue}{\textbf{Blue}}: Depicted is the autoencoder employing a ZINB loss function, using the gene expression matrix as input. In the decoder, the blue rectangle represents the mean of the negative binomial distribution, symbolizing denoised data. The red and green rectangles highlight the two additional parameters of the ZINB distribution, i.e., dispersion and dropout. The associated variables for the gene expression matrix, mean, dispersion, and dropout probabilities are designated as $x$, $\mu$, $\theta$, and $\pi$. \textcolor{pink}{\textbf{Pink}}: Individual representations are aligned with a cluster centroid based on minimal Euclidean distance. The designated cluster centroids function as the cell's discrete representation, with which the autoencoder is then fine-tuned. 
\textcolor{yellow}{\textbf{Yellow}}: This section involves determining the KL divergence among cluster centroids to ascertain their pairs. Within these pairs, representations are categorized as positive examples, while the rest are termed negative samples. The objective of this step is to bolster alignment between the positive example representations by leveraging a contrastive loss in the latent domain.
}
 \label{fig:figure1}
\end{figure}

\noindent \textbf{Autoencoder: }To obtain the initial latent representation of scRNA-seq data, we adopt autoencoder as shown in Fig. \ref{fig:figure1}. Besides, this step can also be viewed as a pretraining step as we will use the trained encoder in the later phase. Autoencoder is an artificial neural network that learns to reconstruct the input data through an encoder and a decoder. The encoder takes the original data as input and compresses the input data to a latent representation which has much lower dimensionality. Then the decoder takes the latent representation as input and learns to reconstruct the original input data from the lower dimensionality. It is characterized by the fact that both the input data and reconstructed output are the same size (i.e. the same number of genes). Although no target information is used, the autoencoder can learn how to effectively compress and subsequently reconstruct the data using the reconstruction loss function in an unsupervised manner. In order to fully reconstruct the data, the autoencoder is forced to learn to compress only the essential information to latent representations and ignore the non-essential information such as random
noise. Therefore, the compressed representation reflects the high dimensional
ambient data space in significantly lower dimensionality and captures the underlying true data manifold. For example, in a data set where snapshots of differentiating blood cells exist, the autoencoder can capture the continuum of differentiation phenotypes in a zero-noise scenario.
As mentioned in \cite{lopez2018deep}, the ZINB model-based autoencoder performed better than models that use alternative distributions in modeling scRNA-seq data, which has a high proportion of zero values and a large variability. ZINB model-based autoencoder uses the likelihood of a ZINB distribution to model scRNA-seq count data. Therefore, unlike the traditional autoencoder model, our model estimates 3 outputs which are mean ($\mu$), dispersion ($\theta$), and dropout ($\pi$) as shown in the decoder part of Fig. \ref{fig:figure1}. Here we formulate the architecture as 
\begin{equation}
\begin{aligned}
E = e_W(X); Z = b_W(E); D = d_W(Z)\\
M^n = exp(DW_\mu^n); \Theta^n = exp(DW_\theta^n); \Pi^n = sigmoid(DW_\pi^n)
\end{aligned}
\end{equation}
where $e_W$, $b_W$, and $d_W$ are the encoder layers, bottleneck layers, and decoder layers with learnable weights $W$, respectively. In this formulation, $X$ is the preprocessed input of the count matrix, where rows and columns correspond to cells and genes. Produced by the encoder, $Z$ is the low-dimensional representation at latent space. Lastly, we define $D$ be the output matrix of the second last hidden layer of the decoder, then for this \textit{naive} autoencoder we append 3 fully connected layers with weights represented by $W_\mu^n$, $W_\theta^n$ and $W_\pi^n$ ($n$ represents they are parameters in \textit{naive} autoencoder) to get the ZINB parameters $M^n$, $\Theta^n$ and $\Pi^n$, which are the matrix form of the estimated mean, dispersion,
and dropout probability, respectively. As the value for mean and dispersion are always non-negative, therefore an exponential output layer is used as the activation function. Since the value for dropout always ranges between zero and one as it represents the probabilities of dropout, the sigmoid layer is used as an activation function.

In our method, ZINB distribution is adopted to fit the reconstructed output. The ZINB distribution is parameterized by a zero inflated negative binomial distribution which consists of the following two components: (1)
a point mass at zero which represents excess zero values in the data and (2) a
negative binomial component representing the count distribution. For scRNA-seq
data, the point mass at zero is parameterized by dropout $\pi$ which represents the probability of dropout events and negative binomial component parameterized by mean $\mu$ and dispersion $\theta$ represents the distribution of the non-zero reads of the count matrix. The formulas are as follows:
\begin{equation}
    NB(x;\mu,\theta) = \frac{\Gamma(x + \theta)}{\Gamma(\theta)}(\frac{\theta}{\theta+\mu})^\theta(\frac{\mu}{\theta+\mu})^x
\end{equation}
\begin{equation}
    ZINB(x;\pi,\mu,\theta) = \pi\delta_0(x) + (1 - \pi)NB(x;\mu,\theta)
\end{equation}

Finally, the reconstruction loss function for our ZINB-based autoencoder is the sum of the negative log of the ZINB likelihood of each data entry:
\begin{equation}
    L_{ZINB}^n = \sum_{ij}-log(ZINB(x_{ij};\pi_{ij}^n,\mu_{ij}^n,\theta_{ij}^n))
\end{equation}
where $i$ and $j$ are the indices of cells and genes, respectively. \\

\noindent \textbf{Vector Quantized Variational Autoencoder: } After the autoencoder is trained in step 1, we first cluster the latent representation using leiden cluster technique. Then the cluster centroids are used as discrete vectors to train another Vector Quantized Variational Autoencoder (VQ-VAE) \cite{van2017neural}. The difference between VQ-VAE and the autoencoder in part 1 is before feeding the latent representation to the decoder, it is first quantized to a set of discrete latent representations to reflect distinct cell types in the dataset. This step is called vector quantization. In this paper, the discrete latent representation corresponds to the cluster centroids we previously mentioned. After the latent representation is quantized, the decoder will take quantized representations as input to reconstruct the gene matrix.

For the detail of quantization, we first create learnable embedding vectors $S$ which are initialized from cluster centroids by taking the average of a cluster. Specifically, the $i$th embedding vector $S_i$ is initialized from the $i$th cluster centroid. We then quantize the output of the encoder $Z$ to one of the cluster centroids based on Nearest Neighbor Search:


\begin{equation}
Z_q=\text { Quantize }\left(Z\right)=S_k \text { where } k=\arg \min _i\left\|Z-S_i\right\|_2
\end{equation}
where $Z_q$ is the quantized latent representation. It actually computes the Euclidean distance between $Z$ and each embedding vector $S$, and then selects the embedding vector $S_k$ with the closest distance as the quantized value $Z_q$. Therefore, the architecture of VQ-VAE can be formulated as:
\begin{equation}
\begin{aligned}
E = e_W(X);
Z = b_W(E);
Z_q=\text { Quantize }(Z);
D = d_W(Z_q)\\
M^q = exp(DW_\mu^q);
\Theta^q = exp(DW_\theta^q);
\Pi^q = sigmoid(DW_\pi^q)
\end{aligned}
\end{equation}


where $W_\mu^q$, $W_\theta^q$ and $W_\pi^q$ are the parameters in VQ-VAE. Besides, both the VQ-VAE and the learnable embedding vector $S$ are trained through the ZINB loss with straight-through estimator \cite{yin2019understanding} to copy the gradient from $Z_q$ to $Z$:
\begin{equation}
    L_{ZINB}^q = \sum_{ij}-log(ZINB(x_{ij};\pi_{ij}^q,\mu_{ij}^q,\theta_{ij}^q))
\end{equation}
Therefore, our VQ-VAE can not only refine the network architecture but also the cluster centroids, which can better identify the contrastive pairs in the next part. \\

\noindent \textbf{Contrastive Learning:} Contrastive learning is exploited to learn robust representations in a self-supervised way \cite{chen2021intriguing}. This part aims to learn effective cell representations with self-supervised contrastive learning by pulling together the representation of functionally similar cells while pushing apart dissimilar cells. Given two single-cell gene expression profiles from two conditions, we constraint the model to produce similar representations for the positive pairs while distinct ones for the negative pairs. By doing so, our encoder can learn a locally smooth nonlinear mapping function that pulls together multiple distortions of a cell in the embedding space and pushes away the other samples. In the transformed space, cells with similar expression patterns form cluster pairs, which are likely to be cells of the same cell types across conditions. To form the positive pairs and negative pairs, we leverage the cluster centroids optimized through the VQ-VAE part, then we assess the similarities of each cluster across two conditions by computing the Kullback-Leibler divergence between the two cluster centroids. The KL divergence is a measure of how one probability distribution diverges from a second, reference probability distribution. KL divergence provides a non-negative value where 0 indicates identical distributions, and a larger value indicates a greater divergence. This measure allows us to quantify the similarity between two clusters in terms of their gene expression profiles, with a lower score indicating higher similarity. We define two cells from two conditions are positive pairs if they are from two clusters with the largest similarity while others are negative pairs. Assume there are two conditions where there are M and N clusters for condition 1 and condition 2, respectively. And, we define positive pair for $n$th cluster $C_{c1}^{n}$ in condition 1 as ${Pair(C_{c1}^{n}) = C_{c2}^{\underset{m \in M}{{\arg\min}} \, KL(C_{c1}^{n}, C_{c2}^{m})} }$ and positive pair for $m$th cluster $C_{c2}^{m}$ in condition 2 as ${Pair(C_{c2}^{m}) = C_{c1}^{\underset{n \in N}{{\arg\min}} \, KL(C_{c2}^{m}, C_{c1}^{n}) }}$. As for cells from the same condition, we define cells from the same cluster are positive pairs while others are negative pairs. The detailed workflows are as follows.

(i) Feature Embedding.  Let $X=\left\{x_k \in R^G\right\}_{k=1}^{B}$ be the queue consisting
of a number of gene expression profiles, where $G$ denotes the
number of genes; $B$ stands for the batch size. In one batch, we first feed $X$ to the encoder to get the latent representation $Z=\left\{z_k \in R^{d_z}\right\}_{k=1}^{B}$, where $d_z$ is number of dimensions of latent representation. Lastly, the latent representations are processed by a projection head $g$ to further map to a $d_h$-dimensional latent space to obtain a new contrastive representation $H=g(Z)=\left\{h_k \in R^{d_h}\right\}_{k=1}^{B}$, where the projection head had been proven that it can facilitate better separation of these positive and negative pairs, which can lead to a more effective contrastive learning process and force the model to learn compact and efficient representations \cite{he2020momentum}.

(ii) Constructing labels. Different from \cite{chen2020simple}, which treats two images that are augmented from the same original image as positive pairs and others are negative pairs. In scBeacon, we consider the positive data points of a contrastive representation $h_i$ as:

\begin{equation}
\begin{split}
H_P^i = \{h_j \in H , |j \neq i, ((\text{Cluster}(h_i) &= \text{Cluster}(h_j)) \wedge (\text{Condition}(h_i) = \text{Condition}(h_j)))\vee \\
((\text{Pair}(\text{Cluster}(h_i)) &= \text{Cluster}(h_j)) \wedge (\text{Condition}(h_i) \neq \text{Condition}(h_j))) \}
\end{split}
\end{equation}

where $Cluster$ and $Condition$ get the cluster and condition of a contrastive representation, respectively. Lastly, we treat the remaining contrastive representation as negative data points:

\begin{equation}
H_N^i = \{h_j \in H , |j \neq i, h_j \notin H_P^i \}
\end{equation}

(iii)  Loss function. 
To distinguish the positive pairs from the negative pairs, we use the following contrastive InfoNCE loss, which can encourage the model to push the positive pairs closer in the embedding space and the negative pairs further apart:

\begin{equation}
L_{con}=-\sum_{i=1}^{N}\log \frac{\sum_{h_p \in H_P^i} e^{\left(h_i \cdot \frac{h_p}{\tau}\right)}}{\sum_{h_n \in H_N^i} e^{\left(h_i \cdot \frac{h_n}{\tau}\right)}} .
\end{equation}
where $\tau$ denotes a temperature parameter. The final loss is computed across all gene expression profiles in a batch. By contrastively learning the gene expression profiles from two conditions from the above InfoNCE loss, our encoder is able to produce similar representations for the positive pairs while making distinct ones for the negative pairs.

\noindent \textbf{Iterative training:} To get better representations from our framework, we train our model iteratively in an end-to-end manner, as shown in Fig. \ref{fig:figure1}.

\vspace{-0.5cm}
\begin{algorithm}
\KwData {Two Single-cell expression matrix from two conditions $X_1, X_2 \in \mathbb R ^{G\times B}$, training parameters.} 
\KwResult {The latent representation for each of the cells}

Train a naive autoencoder with $\left\{ X_1, X_2 \right\}$ with $L_{ZINB}^n$, and get its latent representation $Z$\;

Cluster the representations $Z$ with Leiden clustering\;

\While{The cluster result is not converged}
{Train two VQ-VAE that encoders are initialized from naive autoencoder by treating cluster centroids as discrete embeddings with $L_{ZINB}^q$.\;

Find pairs by maximizing the similarities between clusters.\;

Training the encoder of VQ-VAE with self-supervised contrastive learning by pulling together the representation of functionally similar cells and pushing apart the representation of different cells with $L_{con}$.
}

\caption{Summary of training scBeacon}
\end{algorithm}

\vspace{-0.5cm}
\noindent \textbf{Inference:} After we trained the model, the encoder network of scBeacon is the final productive network, which outputs the representation (a 128-d vector) of a single-cell gene expression profile. After obtaining the representations of all the cells in a dataset, we clustered the cells with the Leiden clustering algorithm. Finally, we find pairs between clusters based on KL divergence between cluster centroids, and then, discover differential expression genes in the cluster.

\vspace{-0.3cm}
\section*{RESULTS}
\vspace{-0.3cm}

\begin{figure*}[!htb]
 \vspace{-1cm}
 \includegraphics[width=\linewidth]{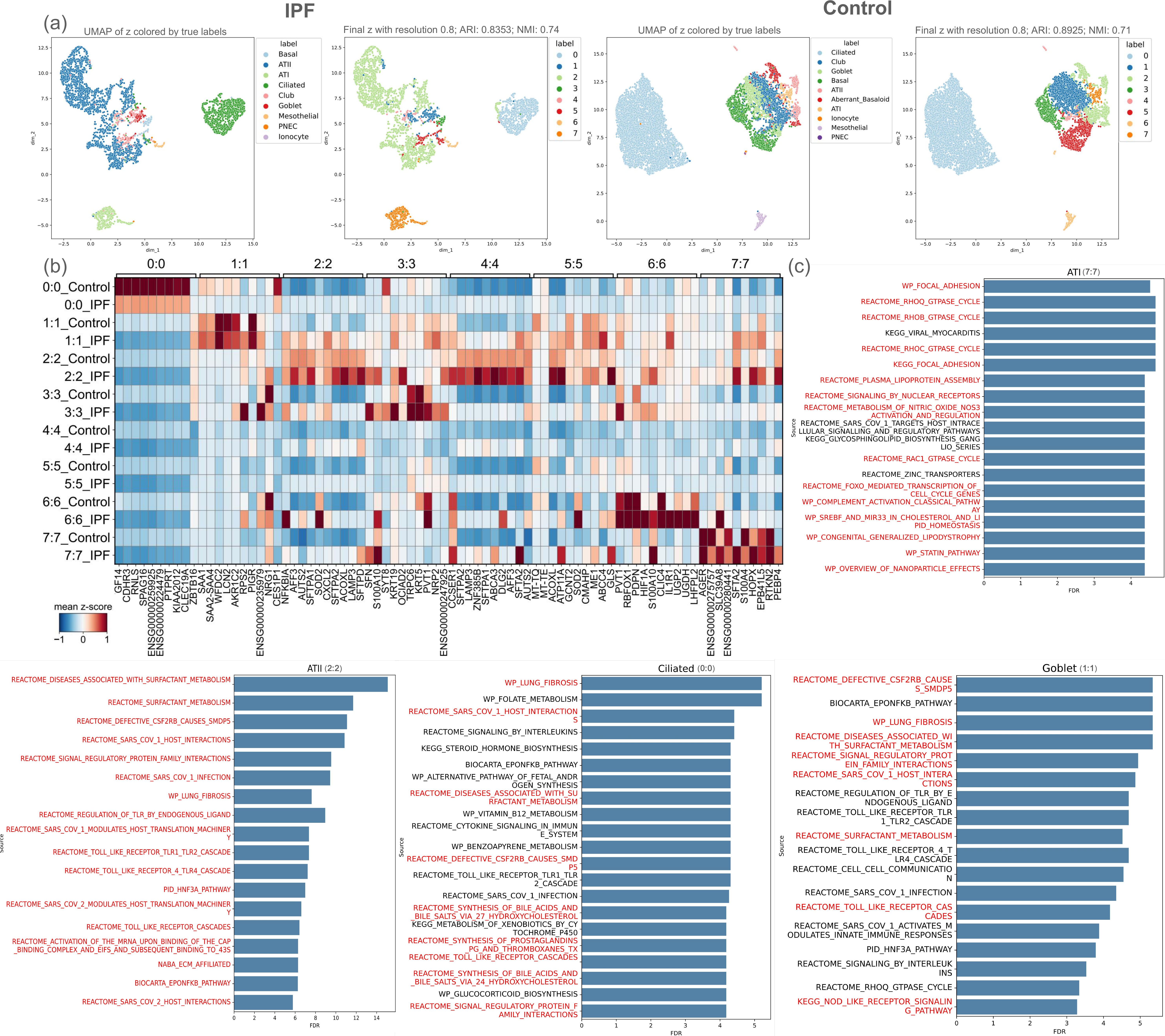}
 \vspace{-0.8cm}
 \caption{\small{\textbf{The scBeacon results on the Atlas Lung dataset}. (a) UMAP visualization of cell embeddings in low-dimensional space, with sequential panels showing cells colored according to verified labels and then segregated by Leiden clustering for IPF and Control samples, respectively. (b) Heatmap detailing the mean expression levels of marker genes for each cluster pair identified. (c) Enriched biological pathways linked to biomarkers pinpointed by scBeacon for four representative cell pairs (cell types or subtypes), with IPF-related pathways distinctly highlighted.
 }}
 \label{fig:dataset1}
 \vspace{-0.6cm}
\end{figure*}

\noindent \textbf{scBeaon effectively identifies biomarkers in the Atlas lung dataset:} 
We commenced our evaluation of scBeacon's capabilities by deploying the method on data derived from the “Atlas of the Lung" dataset as detailed by Adams et al. \cite{adams2020single}. This dataset encompasses samples from both healthy controls and patients diagnosed with idiopathic pulmonary fibrosis (IPF), which are analyzed as representative of two distinct biological conditions. The efficacy of scBeacon in mapping cell cluster pairs and pinpointing biomarkers is demonstrated with distinction in its application for cell embedding learning and downstream clustering analysis. Utilizing UMAP for the visualization of cells in two dimensions, scBeacon displays exceptional skill in grouping similar cells into well-defined clusters and segregating dissimilar ones, as depicted in Fig. \ref{fig:dataset1} (a). The capacity of scBeacon to segregate cell types is not only indicative of its clustering prowess but also of its ability to maintain biological relevancy in the data representation. Moreover, the cluster pairs identified by scBeacon, as shown on the upper x-axis of Fig. \ref{fig:dataset1} (b), show a remarkable correlation with manually annotated cell types. For example, the pairs designated as “0:0", “1:1" and “2:2" align precisely with the manual annotations of “Ciliated", “Goblet" and “ATII" cell types, respectively. This high degree of concordance underscores the precision of scBeacon in cell type identification and its potential for enhancing the accuracy of manual annotations.

Subsequently, a crucial step is the identification of critical marker genes within these clusters that signify the distinction between different biological conditions, such as Healthy control versus IPF. The heatmap in Fig. \ref{fig:dataset1} (b) displays the top-10 markers for each cluster pair, derived from a pool of differentially expressed genes (DEGs). Pathway enrichment analysis, performed using Toppgene \cite{chen2009toppgene}, further validates the effectiveness of the identified markers. As shown in Fig. \ref{fig:dataset1} (c), several pathways significantly enriched between the two conditions (Control vs. IPF) are highlighted in red, marking their direct association with the pathology of IPF. Pathways such as “WP LUNG FIBROSIS," prevalent across epithelial cell types like “ATII", “Ciliated" and “Goblet," underscore the pathological alterations characteristic of IPF, including fibrosis and inflammation \cite{liu2022advances}. The “REACTOME TOLL LIKE RECEPTOR CASCADES," common across these cell types, suggest the potential misfiring of immune responses in IPF \cite{el2019toll}. Furthermore, the “REACTOME DISEASES ASSOCIATED WITH SURFACTANT METABOLISM" pathway, which is also observed in these cell types, implicates abnormal surfactant metabolism in the pathogenesis of lung diseases, including pulmonary fibrosis \cite{whitsett2015diseases}. In the case of the “ATI" cell type, distinct pathways associated with IPF are identified, encompassing the “REACTOME RHO GTPASE CYCLES" like “REACTOME RHOQ", “REACTOME RHOB", “REACTOME RHOC", and “REACTOME RAC1". These pathways are critical to the regulation of actin cytoskeleton remodeling, a key process for fibroblast functionality and myofibroblast differentiation, which are vital to the development of fibrosis in IPF \cite{sward2003role,knipe2015rho,xie2023application}. The systematic comparison with other methods was detailed in the benchmarking section followed.\\

\begin{figure*}[!tb]
 \includegraphics[width=0.9\linewidth,height=15cm]{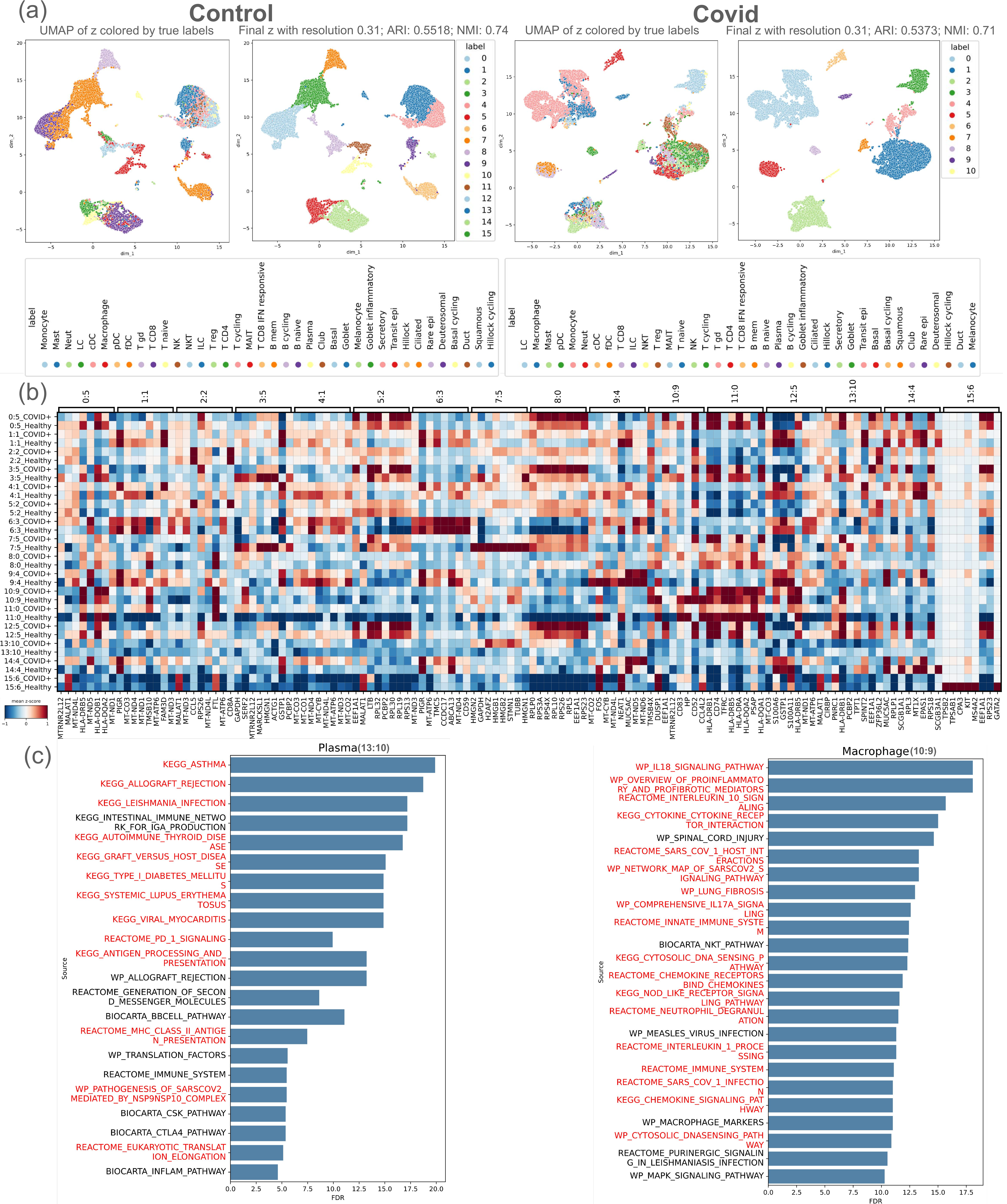}
 \vspace{-0.4cm}
 \caption{\small{\textbf{The scBeacon results on the Covid dataset.} (a) UMAP visualization depicting low-dimensional cell embeddings. Sequential panels display cells colored by actual biological annotations (left) and by Leiden clustering results (right) for Control and COVID-19 samples, respectively. (b) Heatmap of the mean expression levels of selected marker genes across identified cluster pairs, contrasting Control and COVID-19 conditions. (c) Enrichment analysis of pathways for the identified biomarkers, detailing relevant pathways across the two conditions for "Plasma" and "Macrophage" cell types; pathways with known associations to COVID-19 are distinctly highlighted in red.
 }}
 \label{fig:dataset2}
 \vspace{-0.5cm}
\end{figure*}

\noindent \textbf{scBeacon adeptly identifies cluster pairs and biomarkers in the COVID-19 dataset:} We also subjected our method to an evaluation using the COVID-19 dataset from Menon et al. \cite{menon2020sars}, which consists of cell samples from both healthy controls and COVID-19-affected adults. This dataset is notably larger than the Atlas Lung dataset, encompassing a gene expression matrix of impressive dimensions (56488 cells × 33582 genes). The UMAP visualization presented in Fig. \ref{fig:dataset2} (a) may not cluster as distinctly as seen with the Atlas Lung dataset; however, scBeacon still delivers potent clustering results that align accurately with established cell type annotations. For example, the “2:2” cluster pair perfectly maps onto the true “T\_CD8” cell type, while the “6:3” pairing accurately corresponds to the “Ciliated” cell type. Furthermore, the heatmap in Fig. \ref{fig:dataset2} (b) showcases the top-8 genes selected as specific markers for each cell subtype, reflecting a crucial pattern of gene expression across the subtypes. Highlighted among these genes are “HLA-DRB5”, “HLA-DQB1”, and “HLA-DQA2", which are key to the immune response, as their heightened expression in fDC cells has been underscored in recent studies \cite{bernardes2020longitudinal, admati2023two}

The thorough biological pathway enrichment analysis performed on DEGs within “Plasma" and “Macrop-hage" cell types—crucial players in the COVID-19 response \cite{haslbauer2021vascular}—further substantiates scBeacon's capability, as showcased in Fig. \ref{fig:dataset2} (c). This intricate analysis has yielded a set of significantly enriched pathways, underpinning the differential expressions between the examined cohorts and underscoring the precision of scBeacon in pinpointing pivotal genes linked to specific disease mechanisms. Key pathways with direct implications in COVID-19 pathogenesis are highlighted in red for clarity. Among these, the “WP IL18 SIGNALING PATHWAY” operating in “Macrophage" cells stands out, given its vital role in mediating immune responses against viral onslaughts like SARS-CoV-2 \cite{zhang2022cardiovascular, lu2023identification}. Similarly, pathways encompassing “WP OVERVIEW OF PROINFLAMMATORY AND PROFIBROTIC MEDIATORS” relate closely to the inflammatory cascade that characterizes COVID-19 \cite{robb2020non, turnbull2022serum}. Moreover, “KEGG ASTHMA” and “KEGG ALLOGRAFT REJECTION” pathways linked with “Plasma" cells draw a clear connection to COVID-19, with asthma being identified as a risk factor for aggravated COVID-19 symptoms \cite{fang2022bioinformatics} and allograft rejection emerging as a concern post-COVID-19 immunization \cite{rallis2022corneal, vnuvcak2022acute, shah2022acute}. This comprehensive pathway analysis validates the efficacy of scBeacon in biomarker identification and accentuates its clinical significance by bridging the gap between molecular markers and the broader clinical manifestations of COVID-19.\\

\begin{figure}[!tb]
\vspace{-1cm}
 \includegraphics[width=0.9\linewidth]{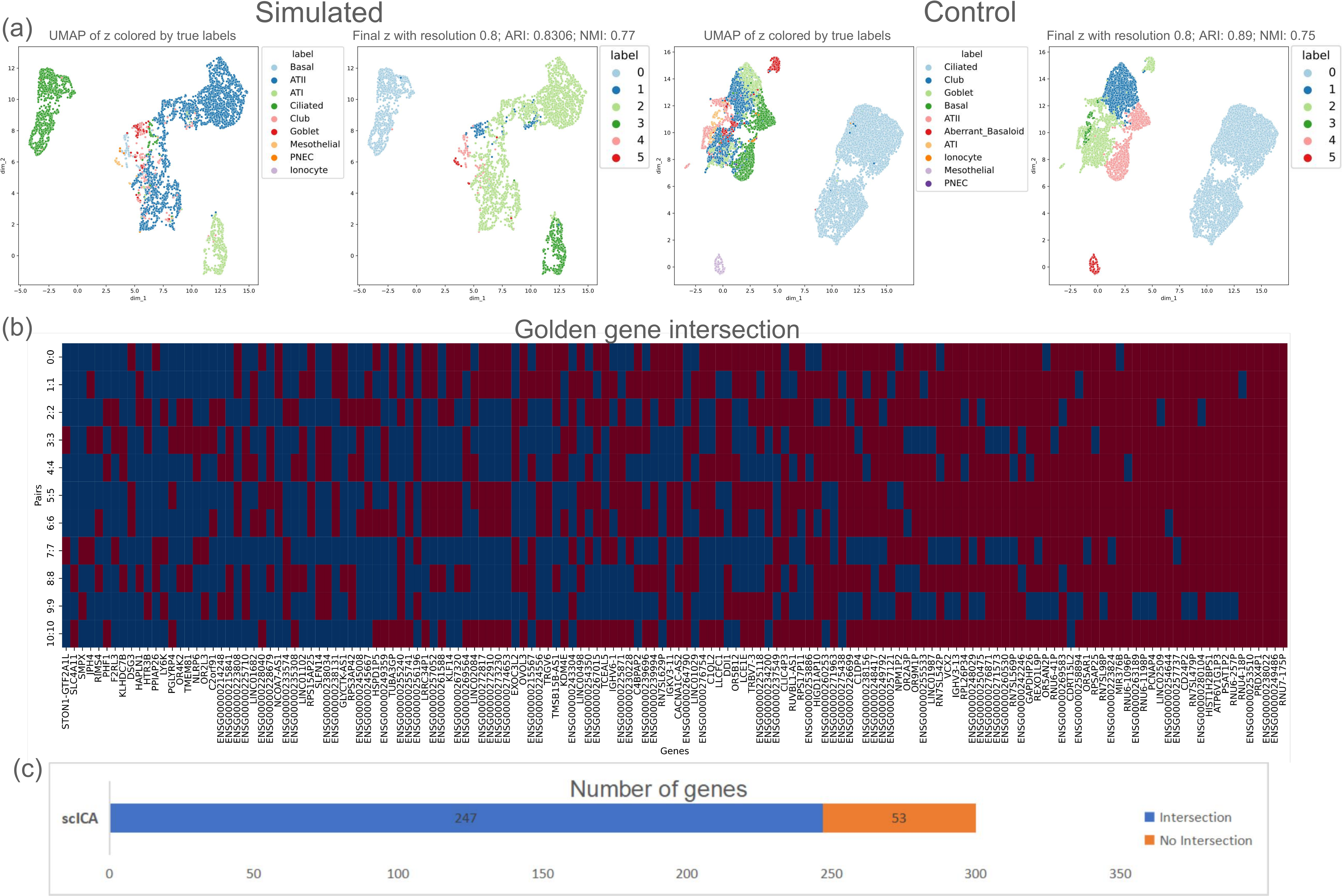}
 \vspace{-0.2cm}
 \caption{\small{\textbf{The results of scBeacon on a Simulated dataset.} (a) UMAP visualization of the low-dimensional representations (colored by leiden clustering and true cell annotations). (b) A heatmap displays the DE genes identified from the golden gene set across various cluster pairs (i.e., cell type or subtypes). The x-axis represents the golden gene set, while the y-axis represents pairs of clusters that have undergone differential gene analysis. (c) A bar plot showing the number of intersections between golden genes and identified DEGs.}}
 \label{fig:dataset3}
 \vspace{-0.5cm}
\end{figure}

\noindent \textbf{The effectiveness of scBeacon is supported by a simulation study.} 
To assess the capability of scBeacon to identify highly expressed genes in controlled conditions, we conducted a simulation study using the Atlas lung dataset as a baseline. The true differentially expressed (DE) genes are not definitively known in actual single-cell datasets, so this simulation offers a controlled environment with ground truth for validation. We amplified the expression values of 300 genes by a factor of five in the Control sample cells to mimic gene expression perturbation, ensuring the same genes were augmented for each cell. The clustering performance of scBeacon on this simulated data was impressive, as shown in Fig. \ref{fig:dataset3} (a), where the representations of various sub-types, such as Basal cells, are distinctly separated, affirming the robustness of the clustering results and setting a strong basis for subsequent differential gene expression analysis. Fig. \ref{fig:dataset3} (b) illustrates a heatmap that presents the DE genes identified by scBeacon against the “golden standard" gene set. The horizontal axis represents the golden standard set, while the vertical axis indicates the cluster pairs compared. The heatmap color coding—with red indicating detected golden standard genes and blue representing undetected ones—demonstrates the method's sensitivity and specificity: different cluster pairs reveal various DEGs that match the golden standard, and some DEGs are consistently identified across multiple clusters. Furthermore, Fig. \ref{fig:dataset3} (c) showcases a barplot that quantifies the intersection of the golden standard genes with the DEGs discovered by scBeacon. The blue bars signify the number of golden standard genes correctly identified as DEGs, while the orange bars denote the genes that were not captured. This visual quantification not only substantiates the precision of scBeacon in detecting true positives but also provides insights into its overall effectiveness in a simulation setting, thereby underscoring the utility of the tool in a ground truth scenario.

\begin{figure*}[!htb]
 \vspace{-0.6cm}
 \centering
 \includegraphics[width=0.9\linewidth,height=10cm]{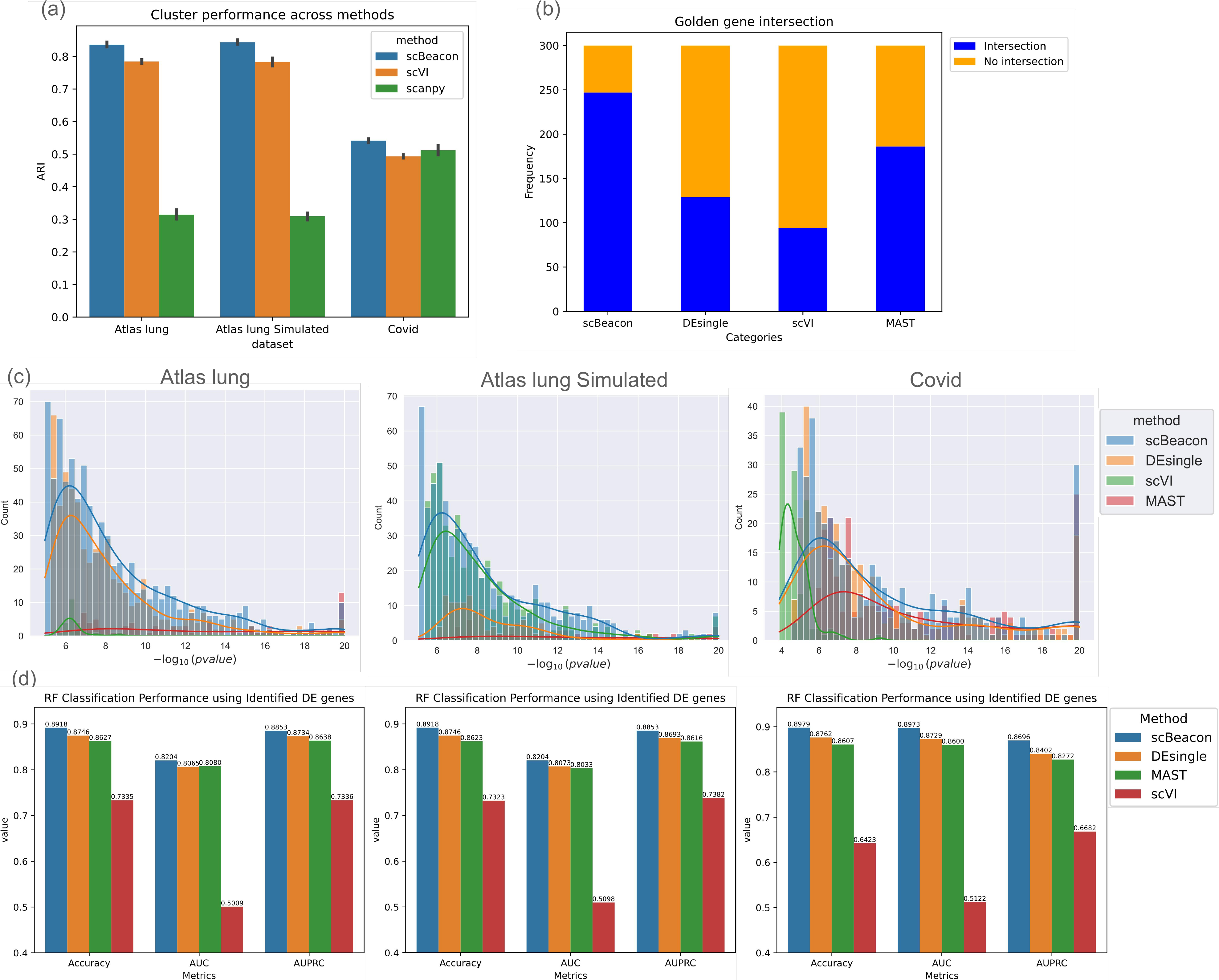}
 \vspace{-0.3cm}
 \caption{\small{\textbf{The benchmark of scBeacon with other methods.} (a) Evaluation of clustering performance using the Adjusted Rand Index (ARI) for various computational methods across different datasets. (b) Overlap of identified “golden standard" genes by various methods within the Atlas Lung Simulated dataset. (c) Density plot illustrating the distribution of p-values for pathway enrichment associated with biomarkers identified for all cell cluster pairs in the dataset, showcasing scBeacon's precision in detecting significant biological pathways. (d) Demonstration of the effectiveness of biomarkers identified by each method based on prediction accuracy of the biological condition, with scBeacon's discovered DEGs contributing to superior prediction performance as reflected by classification metrics (Accuracy, AUC, and AUPRC) using a Random Forest model.
 }}
 \label{fig:benchmark}
 \vspace{-0.6cm}
\end{figure*}

\noindent \textbf{Benchmarking:} We assessed the performance of scBeacon against leading DE analysis methods: MAST, DEsingle, and scVI. MAST utilizes a dual-component generalized linear model designed for the bimodal distribution of scRNA-seq data. DEsingle employs a zero-inflated negative binomial model, suitable for scRNA-seq data, while scVI is a probabilistic approach tailored for scRNA-seq analysis. These methods were benchmarked on clustering performance, DEG identification, and the discriminative power of identified DEGs using a Random Forest classifier. In our comparative analysis, scBeacon demonstrated superior clustering performance, as shown in Fig. \ref{fig:benchmark} (a), with an average increase in the Adjusted Rand Index (ARI) of 5.14, 6.07, and 2.91 points over the second-best method across the three datasets, respectively. Multiple iterations with varied random seeds confirmed the stability of our method, and hyper-parameter fine-tuning for baseline methods ensured fair comparisons. Fig. \ref{fig:benchmark} (b) illustrates scBeacon's modest yet consistent lead in identifying more “golden standard" genes compared to other methods in the simulation study with ground truth, revealing its enhanced detection capabilities. Moreover, panel c presents a density plot comparison of the p-values associated with enriched pathways of identified biomarkers for all cell cluster pairs captured by scBeacon, displaying its adeptness at highlighting significant pathways. The discriminative capability of the identified DEGs was assessed by employing a Random Forest classifier, as detailed in Fig. \ref{fig:benchmark} (d). This involved training and testing the classifier across multiple data partitions, and evaluating model performance in terms of accuracy, AUC, and AUPRC. scBeacon's identified DEGs resulted in better predictive accuracy of biological conditions, outperforming the second-best method on the Atlas Lung dataset, the IPF simulated dataset, and the Covid dataset. Through these comprehensive benchmarks, scBeacon has established its efficacy as a robust tool for clustering, biomarker identification, and elucidating the biological significance of gene expression profiles in single-cell RNA sequencing data analysis.\\
 
\noindent \textbf{Ablation study:} Our ablation study focused on the impact of training iterations and the importance of each component within scBeacon’s architecture, as detailed in Supplementary Fig. S2. The iterative training process on the Atlas Lung dataset revealed improvements in clustering quality, as measured by the Adjusted Rand Index (ARI), which increased steadily up to the tenth iteration where it stabilized, suggesting the efficacy of the method in enhancing latent representations—this trend is depicted on the left of Supplementary Fig. S2 (a). Additionally, when assessing the model's performance without its principal parts—the VQ-VAE and contrastive learning elements—we observed a decline in clustering ARI by 26\% and 24\%, respectively. This underlines the critical role of each component in the model's design. In Supplementary Fig. S2 (b), scBeacon demonstrates its robustness by outclassing the variant models in pinpointing significant differentially expressed genes (p $<$ 0.05), further emphasizing the comprehensive capability of the integrated model for superior biomarker detection.

\section*{Software availability}
\vspace{-0.3cm}
The methodology is encapsulated in scBeacon, our Python-based tool developed with Pytorch, and is available at \url{https://github.com/mcgilldinglab/scBeacon/}. The repository provides comprehensive instructions, thorough documentation, and illustrative examples for users.

\vspace{-0.3cm}
\section*{Acknowledgements}
\vspace{-0.3cm}
This work was funded in part by grants awarded to [JD]. We gratefully acknowledge the support from the Canadian Institutes of Health Research (CIHR) under Grant Nos. PJT-180505; the Funds de recherche du Québec - Santé (FRQS) under Grant Nos. 295298 and 295299; the Natural Sciences and Engineering Research Council of Canada (NSERC) under Grant No. RGPIN2022-04399; and the Meakins-Christie Chair in Respiratory Research.

\printbibliography
\end{document}